\documentclass[12pt,psfig]{article}
\usepackage{graphicx} 
\usepackage{pdfpages}
\usepackage{amsfonts}
 
\begin{document}
\title{LISA Telescope : Phase noise due to pointing jitter}

\author{Jean-Yves Vinet, Nelson Christensen, \\ Nicoleta Dinu-Jaeger, Michel Lintz, 
 Nary Man, Mikha\"el Pichot \footnote{ARTEMIS, Universit\'e C\^ote d'Azur, Observatoire de la C\^ote d'Azur and C.N.R.S., Nice 06304 France}}

\maketitle
\begin{abstract}
In a space based gravitational wave antenna like LISA,  involving long light paths linking distant emitter/receiver spacecrafts, signal detection amounts to measuring the light-distance variations
through a phase change at the receiver. This is why spurious phase fluctuations due to
various mechanical/thermal effects must be carefully studied. We consider
here a possible pointing jitter in the light beam sent from the emitter. We show how
the resulting phase noise depends on the quality of the wavefront
due to the incident beam impinging on the telescope and due to the imperfections of the telescope itself.
Namely, we numerically assess the crossed influence
of various defects (aberrations and astigmatisms), inherent to a  real telescope with pointing fluctuations.  
\end{abstract}
%
\section{Introduction}
It is well-known that ground based Gravitational Wave (GW) antennas like Advanced Virgo~\cite{Acernese} or Advanced LIGO~\cite{Aasi}, by which the  first historical GW signals have been detected~\cite{GW150914,Abbott1,Abbott2}, cannot operate at frequencies below a few Hz due to seismic motions or density fluctuations in the deep ground, which directly couple to the interferometers' mirrors. The very low frequency domain is nevertheless extremely interesting from an astrophysical point of view. This is why from the beginning of the GW detection planning era, as soon as the 1970's, various kinds of space antennas, obviously free of terrestrial issues, have been proposed~\cite{Amaro,Decher}. 

The most recent program is the Laser Interferometer Space Antenna (LISA) proposal, supported by the European Space Agency, in which three spacecrafts orbiting the sun in a triangular constellation exchange light beams propagating along 2.5 Mkm long sides~\cite{CRAS}. The GW signal is expected to be detected through the phase changes at the receiving spacecraft with respect to the local laser. LISA will search for GWs in the $10^{-4}$ to $10^{-1}$ Hz band. Obviously, care must be taken with the various noises able to compete with the extremely small GW signal. 
The LISA optical system will measure the distance between freely falling proof masses in each satellite. In order to detect the GWs it will be necessary to measure the distance between the proof masses with a precision of $< 10$ pm within a measurement bandwidth of 1 Hz~\cite{Livas}.
To maximize the optical coupling, the laser beams are exchanged through an emitting-and-receiving telescope. In this configuration, the amount of scattered light in the optical system is another point to keep under control, as stray light can give rise to noise in the heterodyne phase measurements. 
Because of the large ratio between the emitted and received powers of light through the same telescope it will be necessary to control scattered light~\cite{Amaro,Livas}.

Presented in this paper are our results for the calculations of the phase noise due to pointing jitter for the LISA telescope when the emitted beam contains optical aberrations. This question  has been addressed by Sasso et al. \cite{sassoandco}, who performed numerical evaluations by Monte-Carlo techniques. We present here explicit analytical expressions.
Section \ref{section1} treats the propagation of a laser beam over a Mkm optical path, and shows the compromise between clipping losses (the emitting aperture is finite)  and diffraction losses (the beam expands during propagation).
Section \ref{section2} presents 
the calculation of the phase at the distant telescope when the initial wavefront is affected by distortion and pointing error.

In Section \ref{numerical} the calculations for the laser beam characteristics are carried out in the limit of a weak aberrations approximation, and an important result is derived for the spectral density of the phase noise in terms of a constant (in time) pointing error (bias), a beam jitter, and the characteristics of the measured telescope initial wavefront distortion.
Conclusions are given in Section \ref{conclusion}. 
Section \ref{annex} (annex) 
applies the expression obtained for the phase noise to different asymptotic regimes.
In particular, we  compare our results with those of Sasso et al.~\cite{sassoandco} in the case of a flat emitted beam.
\section{Propagation and clipping trade-off}\label{section1}
We consider a telescope used for sending a 
laser beam at a long distance $L$ (several Mkm) from a circular aperture.  The half aperture (radius) is of the order
of  $a=$ 0.15 m, determined by technical limitations, and the wavelength $\lambda$ is around 1$\mu$m. It is thus clear that we are
in the very far field regime (Fraunhofer) of the diffraction theory.  
The output optical amplitude results essentially
from the truncation by the finite telescope aperture of an ideally Gaussian beam of waist parameter $w$.
The quantity $v\equiv (a/w)^{2}$ is an essential parameter in the present study. 
We assume a complex amplitude of the optical field given by $A(x,y)$, where 
$(x\equiv r\cos\phi,y\equiv r\sin\phi),  (r\le a)$ are the coordinates in the transverse
plane containing the aperture. In the far field, the propagated amplitude $B$  amounts
essentially to the Fourier transform of $A$.
In the following we denote $(p,q)$ the coordinates
in the Fourier space, the Fourier transform of any function $f(x,y)$ is thus $\tilde f(p,q)$,
with 
$$
\tilde f(p,q) \ \equiv \ \int_{\mathbb R^2} e^{ipx+iqy}f(x,y)\,dx\,dy ~ .
$$
With this notation, the far field complex amplitude $B$ propagated from an initial one $A$ from $z=0$, to $z=L$ is :
\begin{equation}\label{BxyL}
B(x,y,L)\ = \ - \frac{i}{\lambda L}\exp\left[ i \, \pi \frac{x^2+y^2}{\lambda L}\right]
\tilde{A}\left( p\equiv \frac{2\pi x}{\lambda L},q\equiv  \frac{2\pi y}{\lambda L}\right) ~ ,
\end{equation}
where now $(x,y)$ are the coordinates in the far plane. 
For an ideal Gaussian beam at waist, truncated at $r=a$, the amplitude at the output of the telescope
would be 
$$
A(x,y)=\sqrt{\frac{2P_0}{\pi w^2}}e^{-r^2/w^2} \  \  (r\le a)  ,
$$ 
where $P_0$ is the laser power, and $w$ the Gaussian waist of the beam.
%
\begin{figure}. 
\centerline{\includegraphics[width=\textwidth]{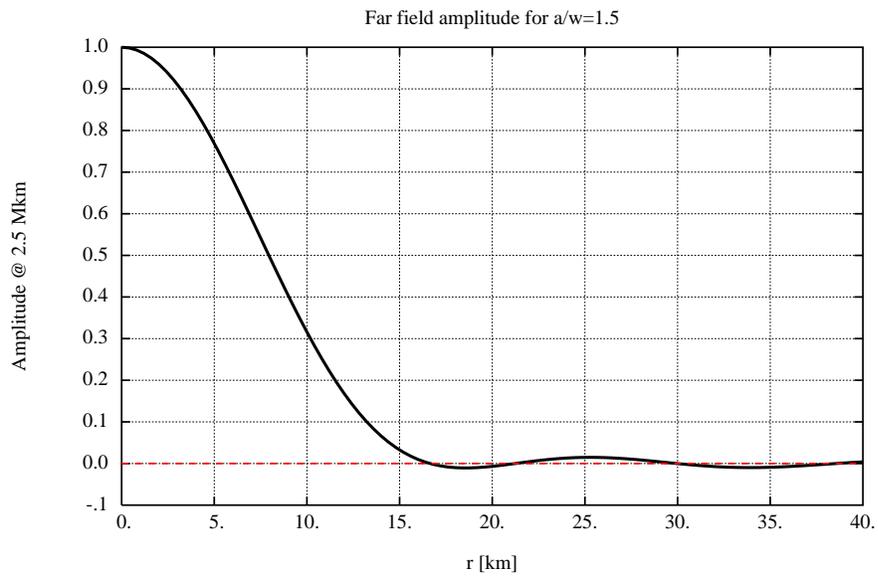}}
\caption{Relative amplitude in the very far field for the ratio $a/w=1.5$ between
the clipping radius $a$ and the waist $w$ at emission. The points at which the amplitude crosses zero (red line) 
correspond to dark rings in the diffraction pattern.}
\end{figure}
\begin{figure}
\centerline{\includegraphics[width=\textwidth]{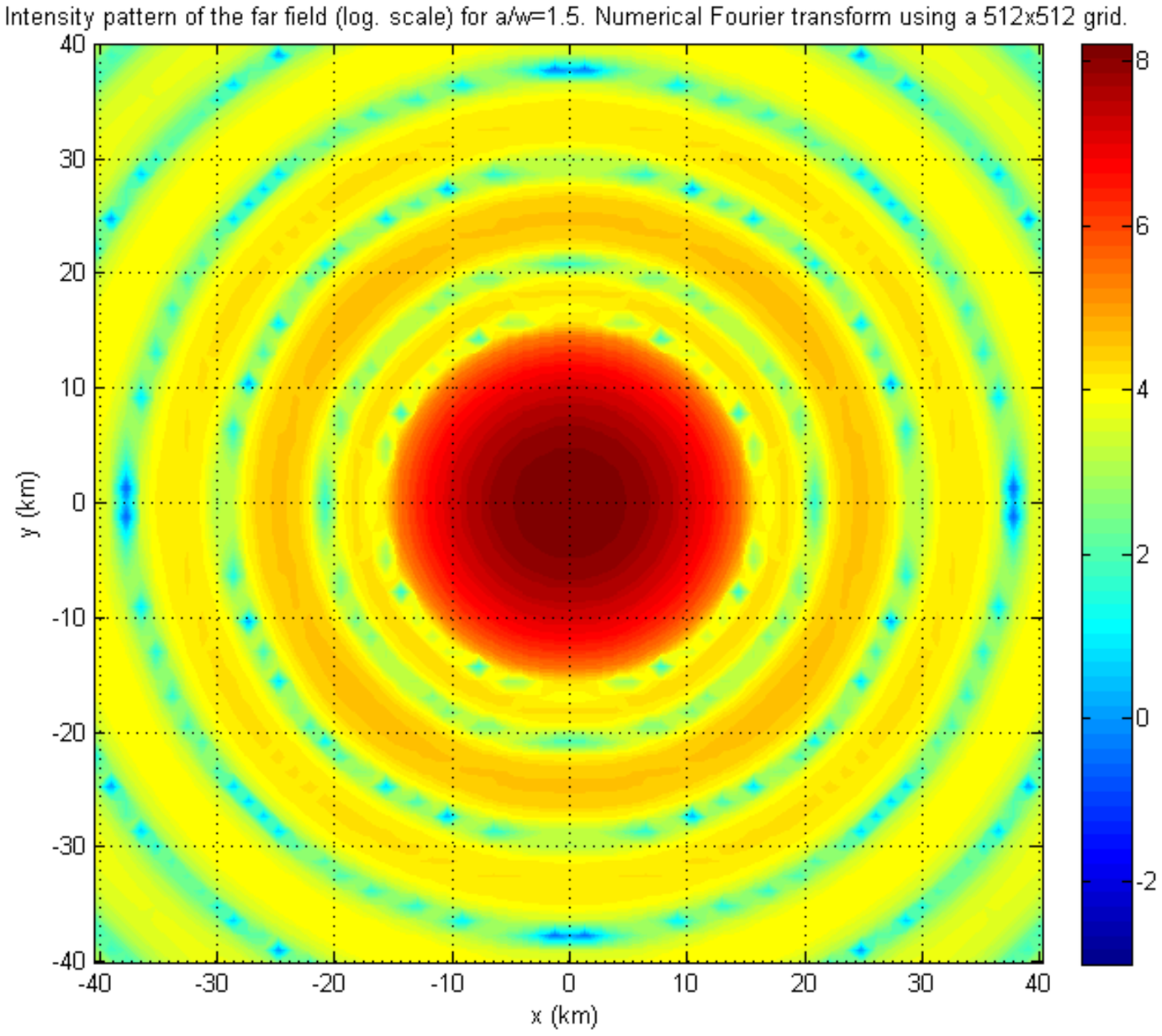}}
\caption{Intensity pattern of the far field (log$_{10}$. scale) for $a/w=1.5$. Units=km. After a numerical Fourier transform
using a 512$\times$512 grid. Arbitrary normalization: we are only interested in the global structure of the field.}
\end{figure}
The Fourier transform $\tilde A(p,q)$  of $A(x,y)$, with the notation $(x=r\cos\phi\ , y=r\sin\phi)$ and 
$p=\rho\cos\psi,\ q=\rho\sin\psi$ ($\rho \equiv \sqrt{p^2+q^2}$), due to the axial symmetry, is a function of $\rho$ only:
$$
\tilde{A}(\rho)  \  = \ \sqrt{\frac{2P_0}{\pi w^2}} \int_0^{2\pi}d\phi
\int_0^a r\,dr\,e^{-r^2/w^2} e^{i\rho r \cos(\phi-\psi)} \ = \
$$ $$
=\ 2\pi \sqrt{\frac{2P_0}{\pi w^2}} \int_0^a e^{-r^2/w^2}J_0(\rho r)r\,dr ~ .
$$
Using the series representing the Bessel function $J_0$, this is
$$
\tilde{A}(\rho)  \  = \ \ 2\pi \sqrt{\frac{2P_0}{\pi w^2}}
\sum_{s=0}^\infty (-)^s \frac{1}{s!^2}\left(\frac{\rho}{2}\right)^{2s}
\int_0^a e^{-r^2/w^2} r^{2s+1}dr ~ ,
$$
or as well (with $\rho=2\pi r/(\lambda L)$, $r$ being now evaluated in the far plane, at $z=L$),
$$ 
\tilde{A}(\rho)  \  = \ \ \pi w^2  \sqrt{\frac{2P_0}{\pi w^2}}
\sum_{s=0}^\infty (-)^s \frac{1}{s!^2}\left(\frac{\rho w}{2}\right)^{2s}
\int_0^{a^2/w^2} e^{-t} t^s dt ~ .
$$ 
With the change of variable $r \equiv \lambda L \rho / (2\pi)$, this can be rewritten as
\begin{equation}\label{farfelu}
\tilde A(r) \ 
= \    \sqrt{2\pi P_0} \,a  \sum_{s=0}^\infty (-)^s \frac{1}{s!}
\left(\frac{\pi a r}{\lambda L}\right)^{2s}  \frac{  \gamma_s(v)}{ v^{s+1/2}} ~ ,
\end{equation}
where we used the notation  $v\equiv a^2/w^2$ and the following definition  for convenience,
\begin{equation}\label{gamma}
\gamma_n(x) \equiv \ 1- e^{-x}\sum_{s=0}^n \frac{x^s}{s!} ~.
\end{equation}
The relation with the classical incomplete gamma function $\gamma(n,x)$ is explicitly
$\gamma_n(x) \equiv \gamma(n+1,x)/n!$.
Displayed in Fig. 1 is the amplitude pattern $\tilde A(r)$ 
at the remote telescope and at distance $r$ from the $z$ axis,
and in Fig. 2 the intensity pattern $|\tilde A(r)|^2$. The diffraction due to clipping at the emission aperture shows a
central lobe of diameter about $\sim$ 30 km.
At a distance $L$,
the amplitude, according to Eq.~(\ref{BxyL}) and Eq.~(\ref{farfelu}) with $r=0$, is a constant over the 
receiving telescope aperture (we take $p=q=\rho=0$) : 
\begin{equation}
B(0,0,L)\ = \ - \frac{i}{\lambda L}\sqrt{\frac{2P_0}{\pi w^2}}
\int_0^{2\pi} d\phi \int_0^a r\,dr\, e^{-r^2/w^2}  
\end{equation}
$$
= \  - \frac{i}{\lambda L}\sqrt{ 2 P_0 \pi w^2} \left(1\ - \ e^{-a^2/w^2} \right) ~ ,
$$
so that the intensity is
\begin{equation}
I(0,0,L) \ = \ \mid B(0,0,L) \mid^2 \ =  \  \frac{2P_0 \pi w^2}{\lambda^2 L^2}
\left(1\ - \ e^{-a^2/w^2} \right)^2 ~ .
\end{equation}
The collected power by the circular aperture of the receiving telescope
of radius $a$ is therefore, assuming a uniform value of the received intensity :
\begin{equation}
P_L\ = \ \frac{2 P_0 \pi^2 a^2 w^2}{\lambda^2 L^2}\left(1\ - \ e^{-a^2/w^2} \right)^2 \ = \
\frac{2 P_0 \pi^2 a^4}{\lambda^2 L^2} \frac{\gamma_0(v)^2}{v} ~ .
\end{equation}
Note that in the definition of $v$ ($v\equiv a^2/w^2$),
the value of $a$ is fixed by technical constraints ($a\sim$ 15 cm is a reasonable value considering the LISA mission), 
so that, through $v$,
it is in fact the beam parameter $w$ that we assume adjustable.
The ratio of received power at length $L$, $P_L$ to the initial laser power $P_0$ can be expressed as
\begin{equation}
\frac{P_L}{P_0} \ = \ 2 (\pi \mathcal F)^2 R(v) ~ ,
\end{equation}
where
\begin{equation}
R(v) \ \equiv \ 
\frac{\left(1-e^{-v}\right)^2}{v}
\end{equation}
and
$\mathcal F \equiv a^2/ (\lambda L)$ is the Fresnel number. For $a=$ 15 cm and $L=$ 2.5 Mkm,
the order of magnitude is $2 (\pi \mathcal F)^2 \sim \ 1.41\times 10^{-9}$, or 1.41 pW/mW.
On the other hand, the clipping of the beam by an aperture of radius $a$ causes a relative loss of power $R_0$
given by 
\begin{equation}
R_0(v) \ = \ 1 - \  \frac{1}{P_0} \int_{\Delta} | A(x,y)|^2 dx\,dy \ = \ e^{-2v} ~ ,
\end{equation}
where $\Delta$ refers to the disk of radius $a$.
\begin{figure}
\centerline{\includegraphics[width=1.0\textwidth]{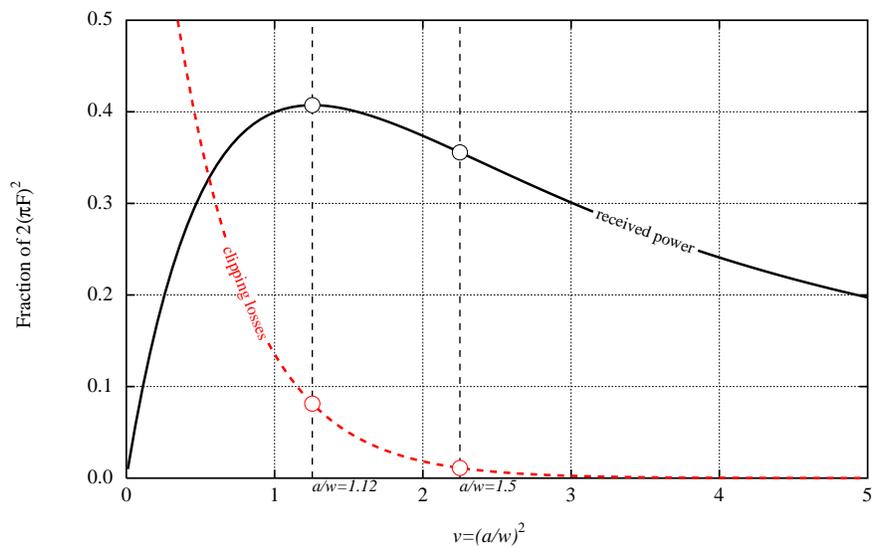}}
\caption{Clipping trade-off. Relative captured power at long distance (percentage
of $2(\pi\mathcal F)^2$): $R(v)$ (black, solid) and corresponding relative truncation losses
at emission: $R_0(v)$ (red, dashed). It can be seen that going from $a/w=1.12$ to $a/w=1.5$
causes a small decrease of captured power, but a strong decrease of clipping losses, i.e. 
of light diffused in the spacecraft structure and expected to cause various issues.}
\end{figure}
For a given value of $a$, a much smaller value of $w$ causes a large angle of diffraction, resulting in a small amount of power captured by a similar aperture (of radius $a$) at long distance. Inversely, a much larger value of $w$ causes by truncation a large loss of power at emission. This is why a trade-off must be considered between these two extreme situations.
If we consider Fig. 3,
we see that the optimum of received power is about  40\%  of the scale factor
$2(\pi \mathcal F)^2$ with 
$a/w \sim 1.12$, (or $v_1 \sim 1.25$)  for about 8\%  power lost at emission by truncation. 
We also see that the choice of  $a/w \sim 1.5$ (or $v_0=2.25$), for instance, 
is not so far from optimal ($R_{0} = 0.35$  instead of $0.4$), and corresponds to a 
much lower fraction of clipped laser power
($\sim 1$ \%). It could be better, regarding scattered light issues,
to have such a reasonable clipping loss. We shall consider in the following the two options.

Let us note that, in a real implementation of the LISA optical metrology system, the telescope essentially produces an image, at the emitting aperture, of the "interface aperture" located on the optical bench (see Sec. 4.4 of ~\cite{Amaro}). The clipping of the beam does not take place at the telescope output aperture or in the telescope structure, but on the optical bench, at the interface aperture. This is indeed where the clipped laser power is to be dumped with high rejection. But the trade off between the clipping factor and the collection efficiency remains the same.
\section{Imperfect emitted amplitude}\label{section2}
The aim of this section is to analytically evaluate the phase noise in the LISA detection system caused
by pointing fluctuations. In absence of constant pointing bias and aberrations, the effect of those fluctuations
would be negligible for an ideal beam : A perfectly spherical wavefront is invariant by a rotation. If
however the beam is both imperfect (aberrations) and has a constant pointing error,  there is a coupling 
between the aberrations and the beam jitter.
We are interested in a small zone containing the receiving telescope, 
of same aperture $2a$, thus for very small values of the Fourier components.
For $x,y\sim a$ and $L\sim$ 2.5 Mkm, we have $(p,q)\sim$ 3.5$\times10^{-4}$
m$^{-1}$.  
We assume the emitting aperture being a disk of radius $a$,  emitting a laser beam of
waist $w$, and having defects expressed by a spurious extra displacement 
$F(x,y)$ [meters] of the wavefront.  The initial amplitude is thus of the form  
\begin{equation}
A(x,y,0)  \  = \  e^{-r^2/w^2} e^{i k F(x,y) }  \  \   (k \equiv 2\pi/\lambda) ~ .
\end{equation}
We will expand the spurious displacement in a series of Zernike functions 
\begin{equation} \label{zernexp}
F(x,y)\ = \ \sum_{n,m}\sigma_{n,m}R_n^{(m)}(r/a)\cos(\sin) m\phi ~ ,
\end{equation}
where (recall our notations : $x=r\cos\phi,y=r\sin\phi $) .
We use the notation $\cos(\sin)$ to recall that Zernike functions have both parities corresponding 
to $\pm m$,  without weighting the formulas.
The $R_n^{(m)}$ are the Zernike polynomials~\cite{BWOLF},
\begin{equation}\label{defzernik}
R_n^{(m)}(\rho) \equiv \sqrt{\frac{2(n+1)}{\pi(1+\delta_{m,0})}} \sum_{s=0}^{(n-m)/2}
(-1)^s \frac{(n-s)!}{s![(n+m)/2-s]![(n-m)/2-s]!} \rho^{n-2s}
\end{equation}
and $\sigma_{n,m}$ (units : m) are
scaling factors to be assessed from measurement, after the completion of the telescope. 
The $\sigma_{n,m}$  have the following definition 
\begin{equation}
\sigma_{n,m} \ \equiv \ \int_\Delta R_n^{(m)}(\rho) F(\rho,\phi) \cos(\sin)(\phi )\rho\,d\rho\,d\phi \ \
(\rho\equiv r/a) ~ ,
\end{equation}
where $\Delta$ is the disk of radius $a$.
The propagated field is therefore 
determined by the Fourier transform 
\begin{equation}\label{FTA}
\tilde A(p,q)\ = \ \int_{\Delta}e^{i px}e^{iqy}e^{-r^2/w^2}e^{ikF(x,y)}dx\,dy \  ~ .
\end{equation}
\subsection{Weak aberrations approximation}
\subsubsection{Pointing fluctuations}
We assume a constant pointing error (bias) $(\theta_0,\psi_0)$ due to a
systematic (small) error, plus a very small time dependent jitter $(\theta_1(t),\psi_1(t))$
caused by various possible mechanical/thermal fluctuations. Both $\theta_0$ and
$\theta_1$ are expected in the nanoradian range. We add an extra
phase factor in the integral Eq.~(\ref{FTA}), of the form
\begin{equation}\label{ffac}
e^{ik x(\theta_0 \cos \psi_0+\theta_1 \cos \psi_1)}e^{iky(\theta_0\sin \psi_0+\theta_1 \sin\psi_1)}
\ =  \ e^{i k r \theta \cos(\phi- \psi)}
\end{equation}
with
\begin{equation}\label{deftheta}
\theta \ \equiv \ \sqrt{\theta_0^2+\theta_1^2+2\theta_0 \theta_1 \cos(\psi_0-\psi_1)}
\end{equation}
and
\begin{equation}\label{defpsi}
\psi\ \equiv \arctan\left[\frac{\theta_0\sin \psi_0+\theta_1\sin\psi_1}{\theta_0\cos\psi_0+\theta_1\cos\psi_1}\right] ~ .
\end{equation}
With these conventions, the far field at the receiver telescope ($p=q=0$) is (with Eq.~(\ref{ffac}))  
\begin{equation}\label{A00}
\tilde{A}(0,0)\ = \  \int_{\Delta}e^{ikr \theta\cos(\phi-\psi)}e^{-r^2/w^2}e^{ikF(x,y)}dx\,dy ~ .
\end{equation}
\subsubsection{First order expansion}
In a realistic device, aiming to send a laser beam over Mkms, one can assume high quality optics, so that  the aberrations
in the optical system are small compared to the wavelength (i.e. $\mid kF\mid \ll1$). We therefore use the
first order approximation (in $kF$) of Eq.~(\ref{A00})
\begin{equation}
\tilde A(0,0) \ = \ \int_{\Delta}e^{ikr \theta\cos(\phi-\psi)} e^{-r^2/w^2}
\left[1+i k F(x,y)\right] dx\,dy
\end{equation}
With Eq.~(\ref{zernexp}), this is
\begin{equation}
\tilde A(0,0) \ = \ \int_0^{2\pi}d\phi\,\int_0^a r\,dr\,e^{ikr \theta\cos(\phi-\psi)} e^{-r^2/w^2}  \ + \
\end{equation}
$$
+ \ ik \sum_{n,m} \sigma_{n,m}  \int_0^{2\pi}d\phi\,\int_0^a r\,dr\,e^{ikr \theta\cos(\phi-\psi)} e^{-r^2/w^2}R_n^{(m)}(r/a) \cos(\sin)m\phi  ,
$$
We may use the well known formula,
\begin{equation}
\int_0^{2\pi} e^{iz\cos\alpha} \cos(\sin)(m(\alpha+\beta))d\alpha \  = \
2\pi i^m J_m(z)\cos(\sin)(m\beta) \  \   (\forall m \in \mathbb N) ~ ,
\end{equation}
where the $J_m(z) \ (m\in \mathbb N$) are the Bessel functions of the 1st kind.
We then get
\begin{equation}
\tilde A(0,0) \ =  2\pi \int_0^a r\,dr\, e^{-r^2/w^2}J_0(kr\theta) \ + \
\end{equation}
$$
+ \ 2 i k\pi \sum_{n,m}\sigma_{n,m}i^m \int_0^a r\,dr\, e^{-r^2/w^2}R_n^{(m)}(r/a) J_m(kr\theta)\cos(\sin)m\psi ~ .
$$
This prompts two remarks.
Firstly, only even order in $m$ will contribute to the phase (pure imaginary terms). A consequence
is that in the preceding series $m$ is even, and consequently, regarding the
structure of Zernike polynomials,  $n$ too.  
The spurious phase of the field is thus, for contributing terms,
\begin{equation}\label{deltaf}
\delta \Phi_{2n,2m} \ = \ k ~ \sigma_{2n,2m} i^{2m}
\frac{   \int_0^a r\,dr\, J_{2m}(k\theta r)R_{2n}^{(2m)}(r/a)e^{-r^2/w^2}}
{\int_0^a r\,dr\,J_0(k\theta r)e^{-r^2/w^2}} \cos(\sin)2m\psi ~ .
\end{equation}
The special case $\delta\Phi_{0,0} \ = \ \frac{1}{\sqrt \pi} k\sigma_{0,0}$ is a constant phase factor (piston) and
has no dependence on $\theta$.
Secondly,  $kr\theta$ being
so small, high orders in $m$ may be neglected (for small $\mid z \mid$, $J_m(z)\sim (z/2)^m/m!$).  If we limit the expansion to the second
order, we have contributions coming from ($2n,0$) and $(2n,2)$. In the following two special
cases, the definition
Eq.~(\ref{defzernik}) may be rewritten as 
\begin{equation}
R_{2n}^{(0)}(\rho)\   \equiv \ (-1)^n \sqrt{\frac{2n+1}{\pi}}\sum_{s=0}^n (-1)^s
 \frac{(n+s)!}{(n-s)! s!^2}\rho^{2s} ~ ,
\end{equation}
and
\begin{equation}
R_{2n}^{(2)}(\rho) \   \equiv \ (-1)^{n-1}  \sqrt{\frac{2(2n+1)}{\pi}}\sum_{s=0}^{n-1} (-1)^s
 \frac{(n+s+1)!}{(n-s-1)! s! (s+2)!}\rho^{2s+2} ~ .
\end{equation}
Substituting  $J_0(z)\sim 1- z^2/4$ and $J_2(z)\sim z^2/8$ in (\ref{deltaf}) yields 
\begin{equation}
\delta\Phi_{2n,0} \ = \ k \sigma_{2n,0}\, \frac{N_n}{D_0}
\end{equation}
where
\begin{equation}
D_0 \ \equiv \ \int_0^a e^{-r^2/w^2}\left( 1- \frac{(k\theta r)^2}{4} \right)\,r\,dr
\  = \  \frac{w^2}{2} \left(\gamma_0(v) - \frac{k^2 a^2 \theta^2}{4 v} \gamma_1(v) \right) ~ .
\end{equation}
We have also :
\begin{equation} \label{bignn}
N_n \ \equiv \ (-1)^n  \frac{w^2}{2}  \sqrt{\frac{2n+1}{\pi}}\sum_{s=0}^n (-1)^s
\frac{(n+s)!}{(n-s)! s!} \frac{1}{v^s} \left[\gamma_s(v)-
(s+1)\frac{k^2 a^2 \theta^2}{4 v}\gamma_{s+1}(v)\right] ~.
\end{equation}
In the same way, we have
\begin{equation}
\delta\Phi_{2n,2} \  = \ k\sigma_{2n,2} \frac{M_n}{D_0}\cos(\sin) 2\psi
\end{equation}
where
\begin{equation}
M_n \equiv \  (-1)^{n-1} i^2  \frac{w^2}{2}  \frac{k^2a^2\theta^2}{8} \sqrt{\frac{2(2n+1)}{\pi}}\sum_{s=0}^{n-1}  (-1)^s
\frac{(n+s+1)!}{(n-s-1)! s!}\frac{1}{v^{s+2}}\gamma_{s+2}(v) ~ .
\end{equation}\label{dphi2n0}
Staying at the second order in $\theta$ finally yields 
\begin{equation}
\delta\Phi_{2n,0} \ = \ k \sigma_{2n,0} \frac{k^2 a^2 \theta^2}{4}(-1)^n\sqrt{\frac{2n+1}{\pi}}
\,\times
\end{equation} 
$$
\times \ \sum_{s=1}^{n}(-1)^s \frac{(n+s)!}{(n-s)!s!}  \frac{ \left[ \gamma_1(v)\gamma_s(v)-
(s+1) \gamma_0(v)\gamma_{s+1}(v) \right] } {v^{s+1}  \gamma_0(v)^2} ~ ,
$$
where a constant term, analogous to a piston was discarded,
and
\begin{equation}\label{dphi2n2}
\delta\Phi_{2n,2} \ = \ k \sigma_{2n,2} \frac{k^2 a^2 \theta^2}{8}(-1)^{n}\sqrt{\frac{2(2n+1)}{\pi}}
\ \times
\end{equation}
$$
\ \times
\sum_{s=0}^{n-1}(-1)^s \frac{(n+s+1)!}{(n-s-1)! s!}\frac{\gamma_{s+2}(v)}{v^{s+2}\gamma_0(v)}
\cos(\sin) 2\psi ~ ,
$$
again with the notation $v \equiv a^2/w^2$.
Eq. 31 and 32  give the results we were looking for : the amplitude of the two main
contributions to the spurious phase due to a jitter of the beam in presence of aberrations. In the following
section, we give numerical results.
\section{Quantitative results}\label{numerical}
\subsection{Phase noise}
We have $k a \theta/2 \ = \ \theta/\Theta$, where $\Theta \equiv \lambda/ (\pi a)$ is of  the order of a
$\mu$rad. The preceding theory thus holds for pointing errors smaller than the beam divergence that do not give rise to a link failure. For a quantitative assessment of the preceding formulas,
we introduce the two following functions ($n \ge 1$),
\begin{equation}\label{deffn}
f_n(v) \equiv \   6 \sum_{s=1}^{n}(-1)^s \frac{(n+s)!}{(n-s)!s!}  \frac{ \left[ \gamma_1(v)\gamma_s(v)-
(s+1) \gamma_0(v)\gamma_{s+1}(v) \right] } {v^{s+1}  \gamma_0(v)^2}
\end{equation}
and
\begin{equation}\label{defgn}
g_n(v) \ \equiv \   3 \sum_{s=0}^{n-1}(-1)^s \frac{(n+s+1)!}{(n-s-1)! s!}\frac{\gamma_{s+2}(v)}{v^{s+2}\gamma_0(v)} ~ ,
\end{equation}
These functions have a simple behavior for the extreme values of $v$, and allow for orders of magnitude approximations.
A version equivalent to Eqs.~(\ref{deffn},\ref{defgn}), but more appropriate for numerical purpose, is,
after some algebra :
$$
f_n(v)\ = \ \frac{6v^n }{(1-e^{-v})^2}\left[ (1-e^{-v})V_n^{(1)}(v)-v e^{-v}V_n^{(0)}(v)\right] ~ ,
$$
\begin{equation}
g_n(v) \ = \  \frac{3v^n} {1-e^{-v}}V_n^{(1)}(v) ~ ,
\end{equation}
using this family of rapidly convergent series  :
\begin{equation}
V_n^{(m)}(x)\equiv  \frac{(n+m)!}{(n-m)!}e^{-x} \sum_{s=0}^\infty \frac{x^s}{s!} \frac{(s+n-m)!}{(s+2n+1)!} \   \
 \ (m \le n) ~ .
\end{equation}
The behavior of functions $f_n(v),g_n(v)$ is shown on Figs.~4 and 5 (the vertical dashed lines correspond to the values $v_0 =2.25$ and $v_1=1.15$).
\begin{figure}\label{f20}
\centerline{\includegraphics[width=1.0\textwidth]{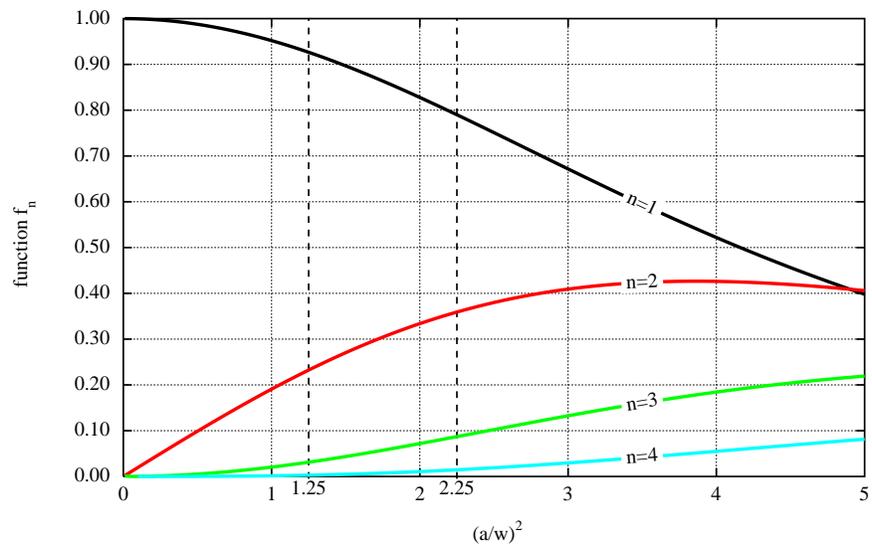}}
\caption{Functions $f_n(v)$, main factors in the magnitude of phase noise due to $(2n,0)$ aberrations.}
\end{figure}
\begin{figure}\label{f22}
\centerline{\includegraphics[width=1.0\textwidth]{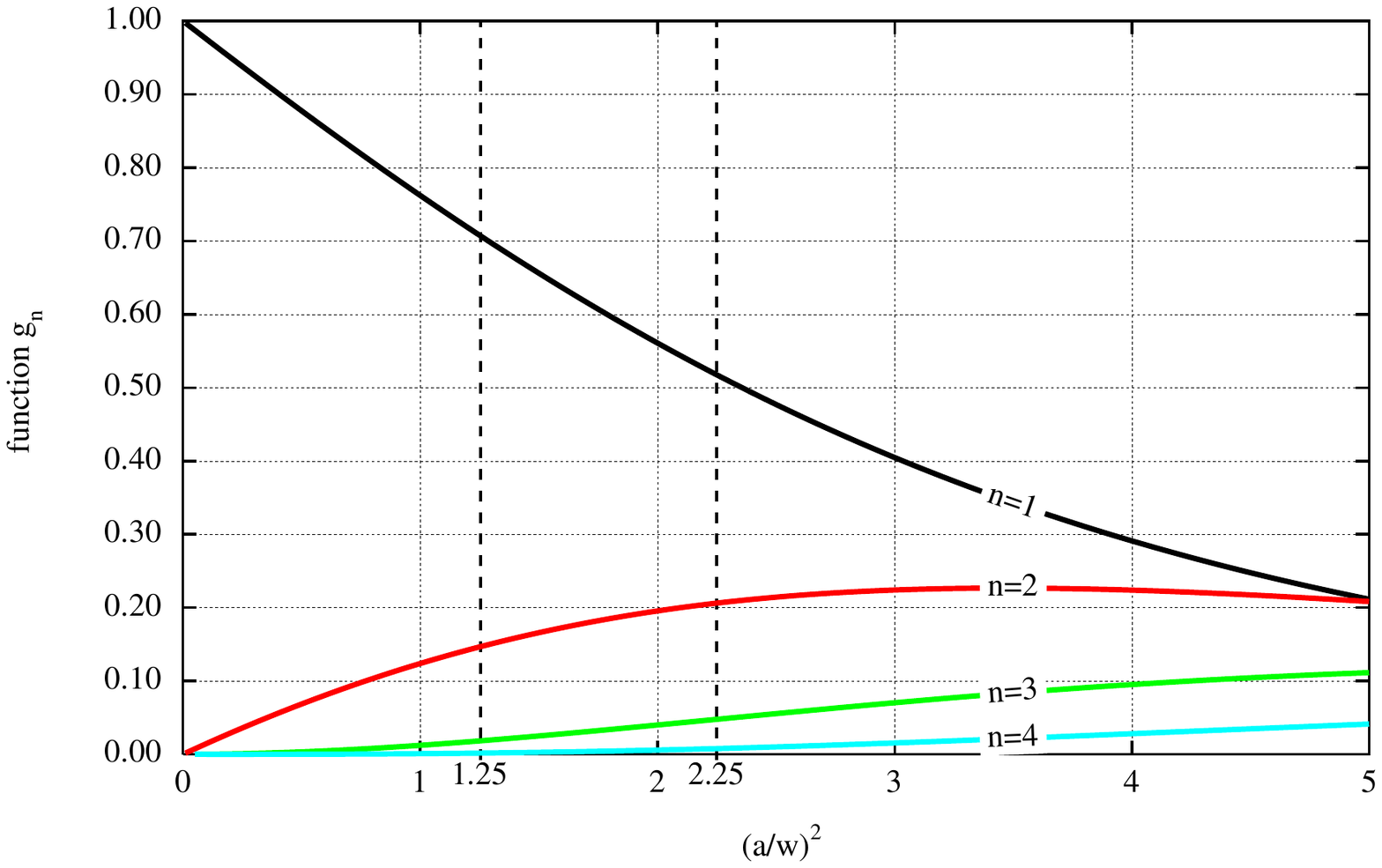}}
\caption{Functions $g_n(v)$, main factors in the magnitude of phase noise due to $(2n,2)$ aberrations.}
\end{figure}
Finally,we have : 
\begin{equation}\label{total}
\delta\Phi(v) \ = \  \sum_{n \ge 1}(-1)^n \left[ \delta\Phi_{2n,0}(v) 
+\delta\Phi_{2n,2}(v)\cos(\sin)2\psi  \right] ~ ,
\end{equation}
with 
\begin{equation}\label{defphinz}
\delta\Phi_{2n,0}= \frac{\sigma_{2n,0} }{\lambda} \frac{k^2 a^2 \theta^2}{12} \alpha_n(v) ~ ,
\end{equation}
\begin{equation}\label{defphin2}
\delta\Phi_{2n,2}= \frac{\sigma_{2n,2}}{\lambda} \frac{k^2 a^2 \theta^2}{12} \beta_n(v) ~ ,
\end{equation}
where the notation
\begin{equation}
\alpha_n(v)\equiv \sqrt{(2n+1)\pi} \, f_n(v), \ \ \beta_n(v)\equiv \sqrt{2(2n+1)\pi}\, g_n(v)
\end{equation}
has been used for brevity.
This normalization choice yields moreover  $f_1(0)=g_1(0)=1$.
After numerical treatment, it appears that these functions have values rapidly decreasing with $n$ for
$v_0 \sim 2.25$ (corresponding to $w=0.1$ m, $a=0.15$ m); see Tables 1 and 2, and Fig. 5.  They decrease
even more rapidly for $v_1=1.25$.
\begin{figure}\label{fngn}
\centerline{\includegraphics[width=1.0\textwidth]{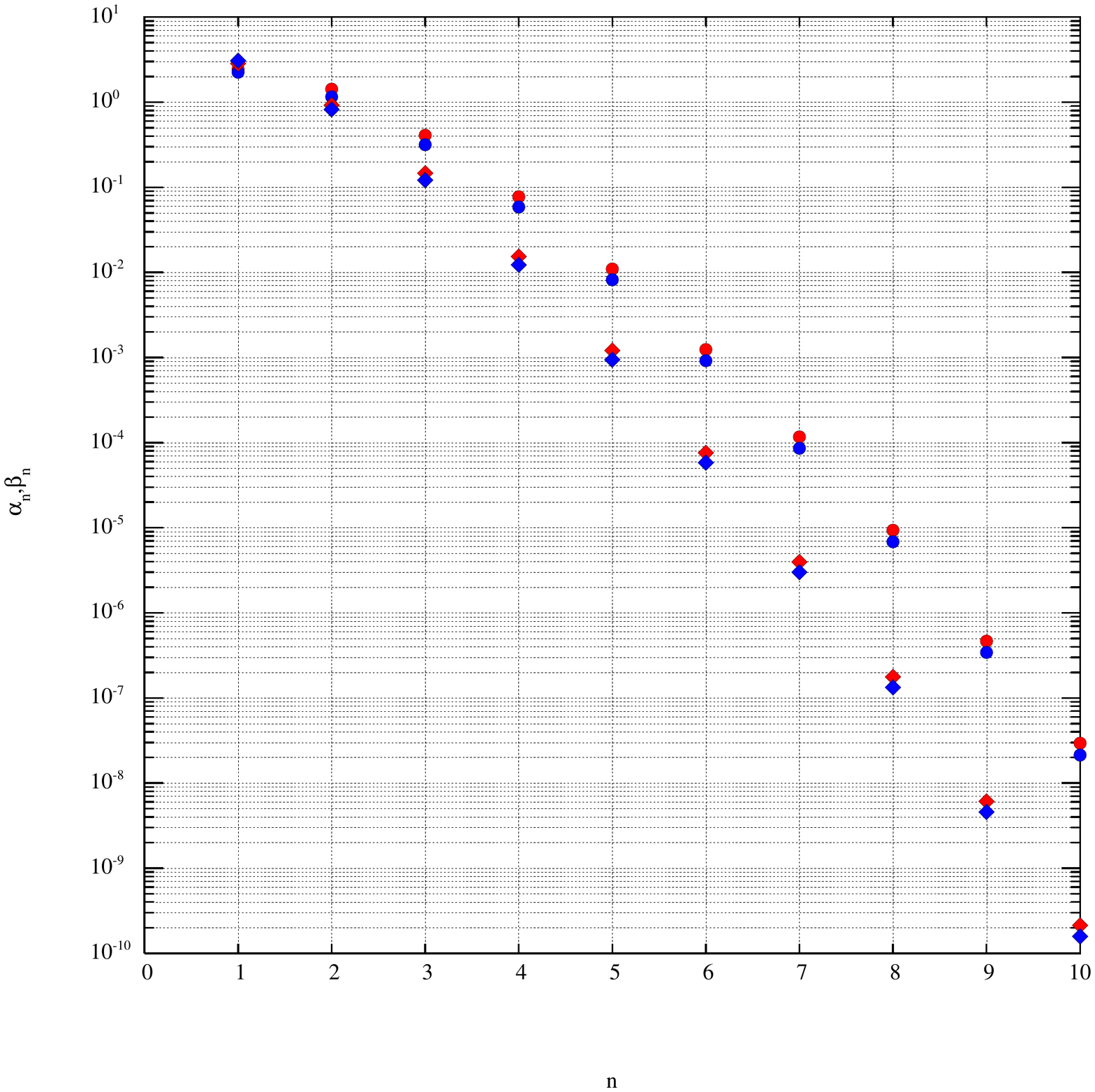}}
\caption{ $\alpha_n(v_0)$  (red circles) and $\beta(v_0)$ (blue circles) for $v_0=$2.25.
$\alpha_n(v_1)$  (red diamonds) and $\beta(v_1)$ (blue diamonds) for $v_1=$1.25}
\end{figure}
\begin{table}[htp]
\caption{$f_n, \ g_n,\ \alpha_n,\ \beta_n$ for $v_0=2.25$}
\begin{center}
\begin{tabular}{|c|c|c|c|c|}
\hline
$n$  &  $f_n(v_0)$  &  $g_n(v_0)$  & $\alpha_n(v_0)$ &  $\beta_n(v_0)$ \\ \hline
1  &  0.790  &  0.518  & 2.425 & 2.247 \\
2  &  0.359  &  0.206 &  1.423 & 1.156 \\
3  &  0.087  &  0.048  & 0.408 & 0.316 \\
4  &  0.015  &  0.008  &  0.077 & 0.059 \\
5 &   0.002  &  9.8\,$10^{-4} $ &  0.011  &  0.008. \\
6 &  1.9 $\times 10^{-4}$  &1.0$\times10^{-4}$  & 1.2$\times 10^{-3}$ &  9.2$\times 10^{-4}$ \\
7 & 1.7 $\times10^{-5}$   &  8.8 $\times\,10^{-6}$ & 1.2 $\times10^{-4}$  & 8.6 $\times10^{-5}$ \\
8 & 1.3 $\times 10^{-6}$ &  6.7 $\times 10^{-7}$   & 9.4$\times 10^{-6}$  & 6.9$\times 10^{-6}$ \\
\hline
\end{tabular}
\end{center}
\label{default}
\end{table}%
\begin{table}[htp]
\caption{$f_n, \ g_n,\ \alpha_n,\ \beta_n$ for $v_1=1.25$}
\begin{center}
\begin{tabular}{|c|c|c|c|c|}
\hline
$n$  &  $f_n(v_1)$  &  $g_n(v_1)$  & $\alpha_n(v_1)$ &  $\beta_n(v_1)$ \\ \hline
1  &  0.926  &  0.708  & 2.844 & 3.073 \\
2  &  0.232  &  0.146 &  0.921 & 0.821 \\
3  &  0.031  &  0.018  & 0.146 & 0.121 \\
4  &  0.003  &  0.002  &  0.015 & 0.012 \\
5 &   2.1$\times10^{-4}$  & 1.1$\times 10^{-4}$  &  0.001  &  0.001 \\
6 &   1.2$\times 10^{-5}$  & 6.4$\times 10^{-6}$  & 7.6$\times 10^{-5}$ &  5.8$\times 10^{-5}$ \\
7 &  5.8$\times 10^{-7}$ & 3.1$\times10^{-7}$ & 4.0$\times 10^{-6}$ & 3.0$\times 10^{-6}$ \\
8 & 2.4$\times 10^{-8} $ & 1.3$\times 10^{-8} $ & 1.8 $\times 10^{-7} $ & 1.3 $\times 10^{-7} $ \\
\hline
\end{tabular}
\end{center}
\end{table}%
\subsubsection{Spectral densities of the phase and length noises}
If we return now to the definition of the angle $\theta$ (see Eq.\ref{deftheta}), 
we consider a jitter angle $\theta_1(t)$ much smaller than the
constant bias $\theta_0$, and moreover if we assume a jitter azimuthal angle
of the form $\psi_1=\psi_{1,0}+\delta \psi_1(t) \  \  (\delta \psi_1(t) \ll \psi_{1,0})$,
then we can write for the time dependent part of $\theta$ 
\begin{equation}
\theta^2(t) \   \simeq \  2\theta_0 \theta_1(t) \cos(\psi_0 - \psi_{1,0}) ~ .
\end{equation}
On the other hand, (Eq.~\ref{defpsi}) gives
\begin{equation}
\psi \ = \ \psi_0 \ + \mathcal O(\theta_1(t)) ~ ,
\end{equation}
so that with Eqs.~(\ref{defphinz}, \ref{defphin2}) we have the following global phase noise 
\begin{equation}
\delta\Phi(t)\ = \ - \sum_{n \ge 1} (-1)^n \delta \Phi_n(t) ~ ,
\end{equation}
with
\begin{equation}
\delta \Phi_n(t) \ = \  \frac{2}{3\Theta^2}  \theta_0 \theta_1(t)
 \left[   \frac{\sigma_{2n,0}}{\lambda} \alpha_n(v) \ + \ 
\,  \frac{\sigma_{2n,2}}{\lambda} \beta_n(v) \cos(\sin)2\psi_0 \right] ~ .
\end{equation}
The linear spectral density of the phase noise is therefore related to the spectral density
of the angular jitter $S_{\theta_1}(f)$  by 
\begin{equation}\label{final}
S_\phi^{1/2}(f) \ =  \   S_{\theta_1} ^{1/2} (f) \frac{2\theta_0}{3\Theta^2}
\sum_{n>0} (-1)^n 
 \left[   \frac{\sigma_{2n,0}}{\lambda} \alpha_n(v) \ + \ 
\,  \frac{\sigma_{2n,2}}{\lambda} \beta_n(v) \cos(\sin)2\psi_0 \right] ~ .
\end{equation}
The spectral density of length noise $S_{\delta L}^{1/2}(f)$  is related to the preceding by
$$
S_{\delta L}^{1/2}(f) \ = \ \frac{\lambda}{2 \pi } \,S_\phi^{1/2}(f) ~ ,
$$
so that finally 
\begin{equation}
S_{\delta L}^{1/2}(f) \ = \  S_{\theta_1} ^{1/2} (f)  \frac{\lambda}{2 \pi} \, \frac{2\theta_0}{3\Theta^2}
\sum_{n>0} (-1)^n 
 \left[   \frac{\sigma_{2n,0}}{\lambda} \alpha_n(v) \ + \ 
\,  \frac{\sigma_{2n,2}}{\lambda} \beta_n(v) \cos(\sin)2\psi_0 \right] ~ .
\end{equation}
With the currently assumed parameters ($a$=0.15m), and with
estimations such as $\theta_0\sim$ 700 nrad~\cite{fitz}, and a spectral
density 
$S_{\theta_1} ^{1/2} (f) \sim 10 {\rm\  nrad/Hz}^{1/2} \times 
\sqrt{  1+({\rm 3mHz} /f)^4}$~\cite{Amaro}, we get the following
order of magnitude expression
\begin{equation}
S_{\delta L}^{1/2}(f) \ = \  1.55\times10^{-10} {\rm m\,Hz^{-1/2}} \sqrt{  1+({\rm 3mHz} /f)^4}\times X ~ ,
\end{equation}
where
$$
X \equiv \sum_{n\ge 1} (-1)^n \left[   \frac{\sigma_{2n,0}}{\lambda} \alpha_n(v_0) \ + \ 
\,  \frac{\sigma_{2n,2}}{\lambda} \beta_n(v_0) \cos(\sin)2\psi_0 \right] ~ .
$$
In the worst case, when all significant aberration terms cumulate (for $n\ge1$), then
a rough order of magnitude is $X\sim8 \sigma /\lambda$ (more precisely 8.13 for $v_0=2.25$, and 7.96 for
$v_1$=1.25), where $\sigma$ is an averaged order of magnitude value of the various 
aberration weights $\sigma_{2n,0},\sigma_{2n,2}$. The result is
\begin{equation}
S_{\delta L}^{1/2}(f) \ \sim \  1 200  {\rm\  pm/Hz}^{1/2}  \sqrt{  1+({\rm 3mHz} /f)^4}\times  \frac{\sigma}{\lambda} ~ .
\end{equation}
This provides a prediction for the acceptable RMS values for the various aberrations if we use the allocation of $\sim$ 2 pm/Hz$^{1/2}$ stated in~\cite{hewitson}.
%
%
\section{Conclusion}\label{conclusion}
A complete far-field modeling of the laser light intensity and phase for LISA is an important task that is not yet complete, but is necessary for the success of the LISA mission. In the study presented in this article we have analytically shown how the deleterious coupling of aberrations in the emitting telescope optics with fluctuating pointing errors (a constant term plus a jitter) may cause an important noise in the optical field collected by the receiving telescope at a distance of 2.5 Mkm. 
 If this noise were to be reduced by special care with the telescope optics, this would imply an RMS wavefront distortion of less than lambda/500, which does not seem to be achievable with state-of-the-art techniques. But two points are worth mentioning. 
On one hand, the limit of 2pm/Hz$^{-1/2}$ considered above is indeed the total allocation for all sources of tilt-to-length coupling (TTL) in the transmit path. The wavefront error considered in this paper is only a fraction of the total TTL, leading to an even more severe limit to wavefront distortions. On the other hand, efforts are being made towards an active control of the position of the aperture on the optical bench, which would allow a major reduction of tilt-to-length coupling~\cite{fitz} and subtraction of the remaining TTL using the Differential Wavefront Sensing signal after the appropriate coefficients have been determined. These aspects are beyond the goal of this paper which is to provide an analytical determination of the coefficients pertaining to the mispointing-induced phase jitter when the emitted wavefront distorsion has been expressed in terms of Zernike polynomials. Our study here, as well as that by Sasso et al. [1], are attempts to address some specific effects, and contribute to the ongoing effort to a comprehensive modeling of
the LISA far-field intensity and phase distributions. Such simulations and modeling are needed to determine the requirements for LISA as it approaches the important period where the design of the full optical system will be defined.
\section{Annex: Two asymptotic regimes }\label{annex}
Eq.(\ref{final}) is much simpler in two opposite limits, corresponding respectively to
very small or very large values of the parameter $v\equiv (a/w)^2$. In these two
cases, it is possible to get the result (phase noise) by 
more straightforward calculations. These checks are useful, in that they help provide confidence in the general result which relies on a somewhat involved calculation.

\subsection{Phase noise with a very large beam waist parameter}
First, let us recall for the reader that $w \gg a$ means that the beam amplitude is uniform across the telescope aperture, so that the emitted beam is mostly an Airy-type beam. Its typical divergence angle is determined by the telescope aperture $2a$.
\subsubsection{Phase noise after the present theory, with $a$ much smaller than $w$}
If the beam waist is large compared to the aperture of the emitting telescope, i.e.  if $v\rightarrow$ 0,
then we see numerically from Eqs.~(\ref{deffn},\ref{defgn}), (see also Figs. 4 and 5), that 
$f_n(0)=g_n(0)=0$ for $n>1$, and $f_1(1)=g_1(1)=1$. Using Eqs.~(\ref{total},\ref{defphinz},
\ref{defphin2}), we get 
\begin{equation}\label{resu311}
\delta\Phi(t) \ = \  - \frac{k^2 a^2 \theta(t)^2}{12}  \sqrt{3\pi} \left[ \frac{\sigma_{2,0}}{\lambda} \ + \
\sqrt{2}\frac{\sigma_{2,2}}{\lambda} \cos(\sin)2\psi \right] ~ .
\end{equation}
\subsubsection{Phase noise : Direct calculation when $a$ is much smaller than $w$}
If the beam amplitude is nearly constant within the aperture of the emitting telescope, we can
take the definition Eq.~(\ref{deltaf}) and ignore the Gaussian factor $e^{-r^2/w^2}$, which
yields 
\begin{equation}
\delta \phi \ = \ k \sum_{n,m} \sigma_{2n,2m}i^{2m} \frac{A_{n,m}}{A_0}\cos(\sin)2m\psi ~ ,
\end{equation}
with
\begin{equation}
A_0 \ =Ê \ \int_0^a J_0(k \theta  r)r dr \ = \ a^2 \frac{J_1(k \theta a)} {k \theta a} ~ ,
\end{equation}
and, using the theorem (see \cite{BWOLF})
\begin{equation}
\int_0^1R_n^{(m)}(\rho) J_m(\rho u) \rho d\rho \  = \ (-1)^{(n-m)/2}\sqrt{\frac{2(2n+1)}{\pi(1+\delta_{m,0})}}
\frac{J_{n+1}(u)}{u} ~ ,
\end{equation}
we also get  
\begin{equation}
A_{n,m} \ = \ \int_0^a J_{2m}(k\theta a) R_{2n}^{(2m)}(r/a)rdr \ = \
(-1)^{n-m} a^2\sqrt{\frac{2(2n+1)}{\pi(1+\delta_{m,0})}} \frac{J_{2n+1}(k \theta a)}{k\theta a} ~ ,
\end{equation}
so that
\begin{equation}
\delta \Phi \ = \ k \sum_{n,m} \sigma_{2n,2m}(-1)^{n-1}i^{2}
\sqrt{\frac{2(2n+1)}{\pi(1+\delta_{m,0})}} \frac{J_{2n+1}(k \theta a)}{J_1(k\theta a} \cos(\sin)2m\psi ~ .
\end{equation}
The $n=0$ term is independent on $\theta$ and may be ignored. Moreover, if we limit
ourselves to second order in $k a \theta$, we are left with $n=1,m=0, 1$, and
eventually, with   $J_3(z)/J_1(z) \sim z^2/24$,
\begin{equation}
\delta \phi \ = -  k \frac{k^2 a^2 \theta^2}{24} \sqrt{\frac{3}{\pi}}  \left[
\sigma_{2,0} +\sqrt{2} \sigma_{2,2} \cos(\sin)2\psi                 \right] ~ ,
\end{equation}
which is identical to Eq.~(\ref{resu311}), and in agreement with the expression
obtained in~\cite{sassoandco} (their Eq.~(24.c)).
\subsection{Phase noise with a very large telescope aperture}
In the case $a \gg w$, the telescope is large enough to emit the Gaussian beam without clipping. The far field is essentially a Gaussian beam with a half-divergence angle $\lambda/\pi w$, and with negligible Airy-type undulations.
\subsubsection{Phase noise after the present theory when $w$ is much smaller than $a$ (almost no clipping)}
If the beam width is small compared to the emitting telescope aperture, i.e. $v\rightarrow\infty$,
then the functions $f_n, g_n$ reduce to :
\begin{equation}
f_n(v) \ = \ 6n(n+1)\,\frac{1}{v^2} + \mathcal O(v^{-4}), \ \, g_n(v) \ = \ 
\ 3 n(n+1)\,\frac{1}{v^2} + \mathcal O(v^{-4}) ~ ,
\end{equation}
giving, with Eqs.~(\ref{total},\ref{defphinz},\ref{defphin2}) :
\begin{equation}\label{resu321}
\delta\Phi(t) \ = \  - \frac{k^2 w^2 \theta(t)^2}{2} \, \frac{w^2}{a^2} \,
\sum_n n(n+1)\sqrt{(2n+1)\pi}\left[ \frac{\sigma_{2n,0}}{\lambda} \ + \
\frac{1}{\sqrt{2}} \, \frac{\sigma_{2n,2}}{\lambda} \cos(\sin) 2\psi \right] ~ .
\end{equation}
\subsubsection{Phase noise : Direct calculation when $w$ is much smaller than $a$}
If $w\ll a$, (i.e. the aperture of the emitting telescope has no clipping effect on the emitted beam) we can:
\begin{itemize}
\item
Neglect the Zernike polynomials of order $m > 2$, and limit others
 at the second order in $r/a$;
\begin{equation}
R_{2n}^{(0)}(r/a)\  = \ (-1)^n\sqrt{\frac{2n+1}{\pi}}\left[1-n(n+1)\frac{r^2}{a^2}\right] +\mathcal O(r^4/a^4) ~ ,
\end{equation}
\begin{equation}
R_{2n}^{(2)}(r/a)\  = \ (-1)^{n-1} \sqrt{\frac{2(2n+1)}{\pi}} \frac{n(n+1)}{2}\frac{r^2}{a^2} +\mathcal O(r^4/a^4) ~ .
\end{equation}
\item
In Eq.~(\ref{deltaf}), replace the limited integration $[0,a]$ by  $[0,\infty]$, so that
the phase change becomes
\begin{equation}\label{item1}
\delta\Phi \  = \   \sum_{n \ge 1} (-1)^n  \frac{A_{n,0}+A_{n,2}\cos(\sin)2\psi}{A_0} ~ ,
\end{equation}
with 
\begin{equation}
A_0 \ = \ \int_0^\infty J_0(k \theta r)e^{-r^2/w^2}r\,dr \ = \ \frac{w^2}{2}e^{-k^2 \theta^2 w^2/4} ~ ,
\end{equation}
\begin{equation}
A_{n,0} = \ \ k \sigma_{2n,0} \,A_0 \sqrt{\frac{2n+1}{\pi}}  \left[1 - n(n+1) \frac{w^2}{a^2} 
\left( 1 - \frac{k^2 \theta^2 w^2}{4}\right) \right] ~ ,
\end{equation}
\begin{equation}
A_{n,2}\ = \ k \sigma_{2n,2}\,A_0 \sqrt{\frac{2(2n+1)}{\pi}} \frac{n(n+1)}{2}\frac{w^2}{a^2} 
\frac{k^2 \theta^2 w^2}{4} ~ ,
\end{equation}
\end{itemize}
so that, for the part depending on $\theta$, we have, according to the definition Eq.~(\ref{item1})
\begin{equation}
\delta\Phi(t) \ = \ - \frac{k^2 w^2 \theta(t)^2}{2} \, \frac{w^2}{a^2} \,
\sum_n n(n+1)\sqrt{(2n+1)\pi}\left[ \frac{\sigma_{2n,0}}{\lambda} \ + \
\frac{1}{\sqrt{2}} \, \frac{\sigma_{2n,2}}{\lambda} \cos(\sin) 2\psi \right] ~ ,
\end{equation}
in agreement with Eq.~(\ref{resu321}).
%
\section{Acknowledgements}
The ARTEMIS Laboratory gratefully acknowledges the support of the Centre National d'\'{E}tudes Spatiales (CNES).

\begin{thebibliography}{99}
%
\bibitem{Acernese}
Advanced Virgo: a second-generation interferometric gravitational wave detector \\
 F Acernese et al.
  Classical and Quantum Gravity, vol.32, n 2, 024001 (2015)
\bibitem{Aasi}
Advanced LIGO \\
  The LIGO Scientific Collaboration  and J Aasi et al. \\
  Classical and Quantum Gravity, vol.32,n 7,074001 (2015)
\bibitem{GW150914}
Observation of Gravitational Waves from a Binary Black Hole Merger \\
Abbott, B. P.  et al. \\
  LIGO Scientific Collaboration and Virgo Collaboration \\
  Phys. Rev. Lett. 116 issue 6, 061102 (Feb. 2016)
\bibitem{Abbott1}
GW170814: A Three-Detector Observation of Gravitational Waves from a Binary Black Hole Coalescence \\
  Abbott, B. P.  et al. \\
  LIGO Scientific Collaboration and Virgo Collaboration \\
  Phys. Rev. Lett. 119 issue 14, 141101.  (Oct. 2017)
\bibitem{Abbott2}
  GW170817: Observation of Gravitational Waves from a Binary Neutron Star Inspiral \\
 Abbott, B. P. et al. \\
 LIGO Scientific Collaboration and Virgo Collaboration \\
 Phys. Rev. Lett. 119 issue 16, 161101 (Oct. 2017)
  \bibitem{Amaro}
  Amaro-Seoane et al. \\
  Laser Interferometer Space Antenna \\
  ArXiv e-prints, eprint 1702.00786  (Feb. 2017)
\bibitem{CRAS}
Some basic principles of a "LISA" \\
Jean-Yves Vinet \\
Comptes Rendus de l'Acad\'emie des Sciences (F), Vol.14 (2013), Issue 4, pp.336-380
\bibitem{Livas}
Optical telescope system-level design considerations for a space-based gravitational wave mission \\
Jeffrey C. Livas,Shannon R. Sankar \\
Proc.SPIE, vol. 9904. (2016)
\bibitem{sassoandco}
   Coupling of wavefront errors and jitter in the LISA interferometer: far-field propagation\\
   C.P. Sasso and G. Mana and S. Mottini \\
  Classical and Quantum Gravity 35, n18, 185013 (2018)
 \bibitem{Decher}
Design Aspects Of A Laser Gravitational Wave Detector In Space \\
Rudolf Decher, Joseph L. Randall, Peter L. Bender, James E. Faller \\
Proc.SPIE, vol.0228. (1980)
\bibitem{BWOLF} M. Born \& E. Wolf \\
Principles of Optics \\
New York, Pergamon, 1989
\bibitem{fitz}
E. Fitzsimons (private communication)
\bibitem{hewitson}
M. Hewitson et al., LISA technical note LISA-LCST-INST-003, "LISA Performance Model and Error Budget", unpublished (2019).
\end{thebibliography}
\end{document}